\documentclass[letter]{aa}
\usepackage{natbib}
\usepackage{graphicx}
\usepackage{txfonts}

\begin{document}

\title{GRB970228 and a class of GRBs with an initial spikelike emission.}

\author{
M.G. Bernardini\inst{1,2}
\and
C.L. Bianco\inst{1,2}
\and
L. Caito\inst{1,2}
\and
M.G. Dainotti\inst{1,2}
\and
R. Guida\inst{1,2}
\and
R. Ruffini\inst{1,2,3}
}

\institute{
ICRANet and ICRA, Piazzale della Repubblica 10, I-65122 Pescara, Italy.
\and
Dipartimento di Fisica, Universit\`a di Roma ``La Sapienza'', Piazzale Aldo Moro 5, I-00185 Roma, Italy. E-mails: maria.bernardini@icra.it, bianco@icra.it, letizia.caito@icra.it, dainotti@icra.it, roberto.guida@icra.it, ruffini@icra.it.
\and
ICRANet, Universit\'e de Nice Sophia Antipolis, Grand Ch\^ateau, BP 2135, 28, avenue de Valrose, 06103 NICE CEDEX 2, France.
}

\titlerunning{GRB970228 and a class of GRBs with an initial spikelike emission.}

\authorrunning{Bernardini et al.}

\date{}

\abstract{
The discovery by \emph{Swift} and HETE-2 of an afterglow emission associated possibly with short GRBs opened the new problematic of their nature and classification. This issue has been further enhanced by the observation of GRB060614 and by a new analysis of the BATSE catalog which led to the identification of a new class of GRBs with ``an occasional softer extended emission lasting tenths of seconds after an initial spikelike emission''.
}{
We plan a twofold task: a) to fit this new class of ``hybrid'' sources within our ``canonical GRB'' scenario, where all GRBs are generated by a ``common engine'' (i.e. the gravitational collapse to a black hole); b) to propose GRB970228 as the prototype of the above mentioned class, since it shares the same morphology and observational features.
}{
We analyze \emph{Beppo}SAX data on GRB970228 within the ``fireshell'' model and we determine the parameters describing the source and the CircumBurst Medium (CBM) needed to reproduce its light curves in the $40$--$700$ keV and $2$--$26$ keV energy bands.
}{
We find that GRB970228 is a ``canonical GRB'', like e.g. GRB050315, with the main peculiarity of a particularly low average density of the CBM $\langle n_{cbm} \rangle \sim 10^{-3}$ particles/cm$^3$. We also simulate the light curve corresponding to a rescaled CBM density profile with $\langle n_{cbm} \rangle=1$ particle/cm$^3$. From such a comparison it follows that the total time-integrated luminosity is a faithful indicator of the nature of GRBs, contrary to the peak luminosity which is merely a function of the CBM density.
}{
We call attention on discriminating the short GRBs between the ``genuine'' and the ``fake'' ones. The ``genuine'' ones are intrinsically short, with baryon loading $B \la 10^{-5}$, as stated in our original classification. The ``fake'' ones, characterized by an initial spikelike emission followed by an extended emission lasting tenths of seconds, have a baryon loading $10^{-4} \la B \leq 10^{-2}$. They are observed as such only due to an underdense CBM consistent with a galactic halo environment which deflates the afterglow intensity.
}


\maketitle

\section{Introduction}\label{intro}

GRB060614 \citep{ge06,ma07} has enlightened the \emph{Swift} ``short GRBs' revolution'' \citep[see e.g.][]{ge05}, also confirmed in GRB050709 \citep{villasenor} by HETE-2. Stringent upper limits on the luminosity of the Supernova possibly associated with GRB060614 have been established \citep{da06}. These sources motivated \citet{nb06} to reanalyze the BATSE catalog identifying a new GRB class with ``an occasional softer extended emission lasting tenths of seconds after an initial spikelike emission''. Such observations challenge the standard GRBs scheme \citep{k92,da92} in which the gamma events originate from ``Hypernovae'' and are branched into two classes: ``short'' GRBs (lasting less than $\sim 2$ s) and ``long'' GRBs (lasting more than $\sim 2$ s up to $\sim 1000$ s). \citet{nb06} suggested that such a scheme ``is at best misleading''. GRB060614, indeed, ``reveals a first short, hard-spectrum episode of emission (lasting $5$ s) followed by an extended and somewhat softer episode (lasting $\sim 100$ s)'': a ``two-component emission structure'' which is ``similar'' to the one analyzed by \citet{nb06}. GRB060614 appears as a long burst, lasting more than $\sim 100$ s, but shares some spectral properties with the short ones as the absence of a spectral lag \citep{ge06}. The authors conclude that ``it is difficult to determine unambiguously which category GRB060614 falls into'' and then GRB060614, due to its ``hybrid'' observational properties, ``opens the door on a new GRB classification scheme that straddles both long and short bursts'' \citep{ge06}.

The aim of this Letter is twofold: a) to fit this new class of ``hybrid'' sources within our ``canonical GRB'' scenario \citep{rlet1,rlet2,XIIBSGC}, where all GRBs are generated by a ``common engine'' (i.e. the gravitational collapse to a black hole); b) to propose GRB970228 \citep{costa,fr98,fr00} as the prototype of the above mentioned class, since it shares the same morphology and observational features.

We define a ``canonical GRB'' light curve with two different components. The first one is the Proper-GRB (P-GRB, a precursor in the GRB) and the second one is the afterglow \citep{rlet2,XIIBSGC}. The P-GRB is emitted at the transparency of the relativistically expanding optically thick ``fireshell'' created in the black hole formation. The afterglow originates from the decelerating optically thin ``fireshell'' of ultrarelativistic baryons interacting with the CircumBurst Medium (CBM).

The ratio between the total time-integrated luminosity of the P-GRB and the corresponding one of the afterglow is the crucial quantity for the identification of GRBs' nature. When the P-GRB is the leading contribution to the emission and the afterglow is negligible we have a ``genuine'' short GRB \citep{rlet2}. In the other GRBs the afterglow contribution is generally predominant. Still, this case presents two distinct possibilities: the afterglow peak luminosity can be either larger or smaller than the P-GRB one. As we show in this Letter, the simultaneous occurrence of an afterglow with total time-integrated luminosity larger than the P-GRB one, but with a smaller peak luminosity, is indeed explainable in terms of a peculiarly small average value of the CBM density and not due to the intrinsic nature of the source. In this sense, GRBs belonging to this class are only ``fake'' short GRBs. We show as a very clear example the case of GRB970228. We identify the initial spikelike emission with the P-GRB, and the late soft bump with the peak of the afterglow. GRB970228 shares the same morphology and observational features with the sources analyzed by \citet{nb06} as well as with e.g. GRB050709 \citep{villasenor}, GRB050724 \citep{campana_short} and GRB060614 \citep[][Caito et al. in preparation]{ge06}. Therefore, we propose GRB970228 as a prototype for this new GRB class.

In sec. \ref{fireshell} we briefly outline our theoretical model. In sec. \ref{observ} we recall GRB970228 observational properties. In sec. \ref{theo} we present the theoretical light curves compared with the GRB970228 prompt emission observations in the \emph{Beppo}SAX GRBM ($40$--$700$ keV) and WFC ($2$--$26$ keV) energy bands. In sec. \ref{rescale} we present some explicit examples in order to probe the crucial role of the average CBM density in explaining GRB970228 observational features. In sec. \ref{concl} we present our conclusions.

\section{The ``fireshell'' model and the Amati relation}\label{fireshell}

We assume that all GRBs, including the ``short'' ones, originate from the gravitational collapse to a black hole \citep{rlet2,XIIBSGC}. The $e^\pm$ plasma created in the process of the black hole formation expands as a spherically symmetric ``fireshell'' with a constant width on the order of $\sim 10^8$ cm in the laboratory frame, i.e. the frame in which the black hole is at rest. We have only two free parameters characterizing the source, namely the total energy $E_{e^\pm}^{tot}$ of the $e^\pm$ plasma and its baryon loading $B\equiv M_Bc^2/E_{e^\pm}^{tot}$, where $M_B$ is the total baryons' mass \citep{rswx00}. They fully determine the optically thick acceleration phase of the fireshell, which lasts until the transparency condition is reached and the P-GRB is emitted \citep{rlet2}. The afterglow emission then starts due to the collision between the remaining optically thin fireshell and the CBM and it clearly depends on the parameters describing the effective CBM distribution: its density $n_{cbm}$ and the ratio ${\cal R}\equiv A_{eff}/A_{vis}$ between the effective emitting area of the fireshell $A_{eff}$ and its total visible area $A_{vis}$ \citep{rlet02,spectr1,fil}. The afterglow luminosity consists of a rising branch, a peak, and a decaying tail \citep{rlet2}.

Therefore, as we recalled in the introduction, unlike treatments in the current literature \citep[e.g.][and references therein]{p04,m06}, within our model we define a ``canonical GRB'' light curve with two sharply different components: the P-GRB and the afterglow \citep{rlet2,XIIBSGC}. The P-GRB has the imprint of the black hole formation, an harder spectrum and no spectral lag \citep{brx01,rfvx05}. The afterglow phase presents a clear hard-to-soft behavior \citep{031203,spectr1,050315}. The peak of the afterglow contributes to what is usually called the ``prompt emission'' \citep[e.g.][]{rlet2,050315,060218}. The ratio between the total time-integrated luminosities of the P-GRB and of the afterglow (namely, their total energies) as well as the temporal separation between their peaks are functions of the $B$ parameter \citep{rlet2}. The ``genuine'' short GRBs correspond to the cases where $B \la 10^{-5}$ (see Fig.~\ref{figX}). In this case the leading contribution to the prompt emission is the P-GRB, prominent with respect to the afterglow. In the limit $B \to 0$ the afterglow vanishes (see Fig.~\ref{figX}). In the opposite limit, for $10^{-4} \la B \leq 10^{-2}$ the afterglow component is predominant and this is indeed the case of most of the GRBs we have recently examined, including GRB970228 (see Fig.~\ref{figX}). For the existence of the upper limit $B \leq 10^{-2}$ see \citet{rswx00} and \citet{060218}.

We turn now to the Amati relation \citep{aa02,amati06} between the isotropic equivalent energy emitted in the prompt emission $E_{iso}$ and the peak energy of the corresponding time-integrated spectrum $E_p$. It clearly follows from our treatment \citep{031203,spectr1,050315} that both the hard-to-soft behavior and the Amati relation occurs uniquely in the afterglow phase which, in our model, encompass as well the prompt emission. The observations that the initial spikelike emission in the above mentioned ``fake'' short GRBs, which we identify with the P-GRBs, as well as all ``genuine'' short GRBs do not fulfill the Amati relation \citep[see][]{amati06} is indeed a confirmation of our theoretical model. We look forward to verifications in additional sources.

\section{GRB970228 observational properties}\label{observ}

GRB970228 was detected by the Gamma-Ray Burst Monitor (GRBM, $40$--$700$ keV) and Wide Field Cameras (WFC, $2$--$26$ keV) on board \emph{Beppo}SAX on February $28.123620$ UT \citep{fr98}. The burst prompt emission is characterized by an initial $5$ s strong pulse followed, after $30$ s, by a set of three additional pulses of decreasing intensity \citep{fr98}. Eight hours after the initial detection, the NFIs on board \emph{Beppo}SAX were pointed at the burst location for a first target of opportunity observation and a new X-ray source was detected in the GRB error box: this is the first ``afterglow'' ever detected \citep{costa}. A fading optical transient has been identified in a position consistent with the X-ray transient \citep{vp}, coincident with a faint galaxy with redshift $z=0.695$ \citep{bloom01}. Further observations by the Hubble Space Telescope clearly showed that the optical counterpart was located in the outskirts of a late-type galaxy with an irregular morphology \citep{sau97}.

The \emph{Beppo}SAX observations of GRB970228 prompt emission revealed a discontinuity in the spectral index between the end of the first pulse and the beginning of the three additional ones \citep{costa,fr98,fr00}. The spectrum during the first $3$ s of the second pulse is significantly harder than during the last part of the first pulse \citep{fr98,fr00}, while the spectrum of the last three pulses appear to be consistent with the late X-ray afterglow \citep{fr98,fr00}. This was soon recognized by \citet{fr98,fr00} as pointing to an emission mechanism producing the X-ray afterglow already taking place after the first pulse.

\section{The analysis of GRB970228 prompt emission}\label{theo}

\begin{figure}
\includegraphics[width=\hsize,clip]{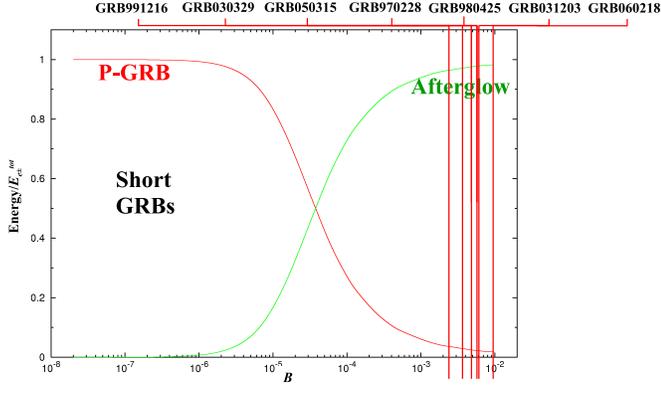}
\caption{The energy radiated in the P-GRB and in the afterglow, in units of the total energy of the plasma ($E_{e^\pm}^{tot}$), are plotted as functions of the $B$ parameter. Also represented are the values of the $B$ parameter computed for GRB991216, GRB030329, GRB980425, GRB970228, GRB050315, GRB031203, GRB060218. Remarkably, they are consistently smaller than, or equal to in the special case of GRB060218, the absolute upper limit $B \protect\la 10^{-2}$ established in \citet{rswx00}. The ``genuine'' short GRBs have a P-GRB predominant over the afterglow: they occur for $B \protect\la 10^{-5}$ \citep{rlet2}.}
\label{figX}
\end{figure}

\begin{figure}
\includegraphics[width=\hsize,clip]{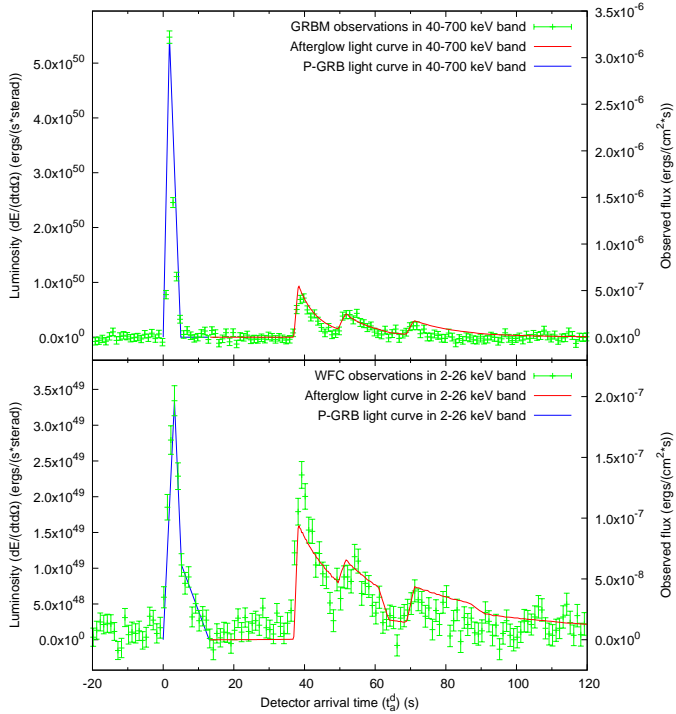}
\caption{\emph{Beppo}SAX GRBM ($40$--$700$ keV, above) and WFC ($2$--$26$ keV, below) light curves (green points) compared with the theoretical ones (red lines). The onset of the afterglow coincides with the end of the P-GRB (represented qualitatively by the blue lines).}
\label{970228_fit_prompt}
\end{figure}

\begin{figure}
\includegraphics[width=\hsize,clip]{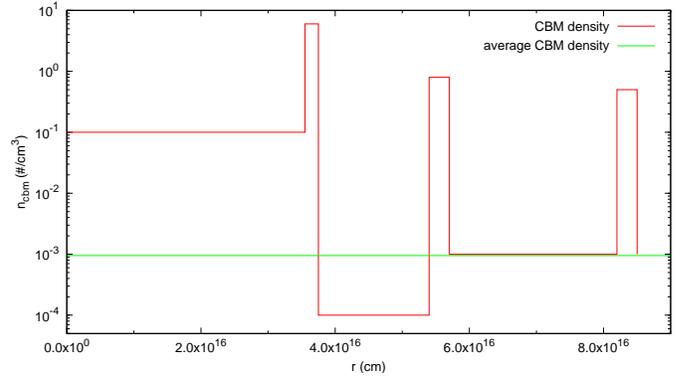}
\caption{The CBM density profile we assumed to reproduce the last three pulses of the GRB970228 prompt emission (red line), together with its average value $\langle n_{cbm} \rangle = 9.5\times 10^{-4}$ particles/cm$^3$ (green line).}
\label{mask}
\end{figure}

\begin{figure}
\includegraphics[width=\hsize,clip]{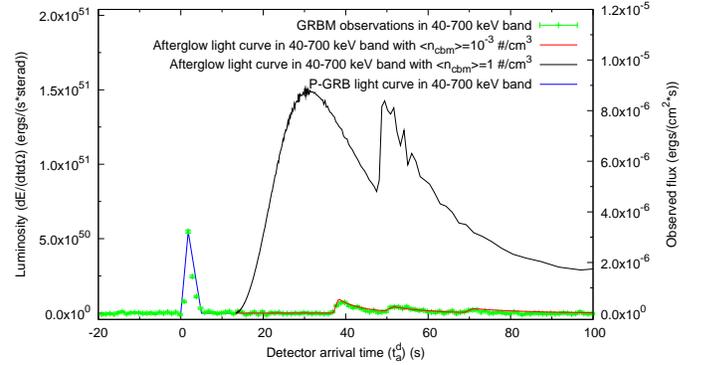}
\caption{The theoretical fit of the \emph{Beppo}SAX GRBM observations (red line, see Fig. \ref{970228_fit_prompt}) is compared with the afterglow light curve in the $40$--$700$ keV energy band obtained rescaling the CBM density to $\langle n_{cbm} \rangle = 1$ particle/cm$^3$ keeping constant its shape and the values of the fundamental parameters of the theory $E_{e^\pm}^{tot}$ and $B$ (black line). The P-GRB duration and luminosity (blue line), depending only on $E_{e^\pm}^{tot}$ and $B$, are not affected by this process of rescaling the CBM density (see sec. \ref{fireshell}).}
\label{picco_n=1}
\end{figure}

In Fig.~\ref{970228_fit_prompt} we present the theoretical fit of \emph{Beppo}SAX GRBM ($40$--$700$ keV) and WFC ($2$--$26$ keV) light curves of GRB970228 prompt emission \citep{fr98}. Within our ``canonical GRB'' scenario we identify the first main pulse with the P-GRB and the three additional pulses with the afterglow peak emission, consistently with the above mentioned observations by \citet{costa} and \citet{fr98}. Such last three pulses have been reproduced assuming three overdense spherical CBM regions (see Fig.~\ref{mask}) with a very good agreement (see Fig.~\ref{970228_fit_prompt}).

We therefore obtain for the two parameters characterizing the source in our model $E_{e^\pm}^{tot}=1.45\times 10^{54}$ erg and $B = 5.0\times 10^{-3}$. This implies an initial $e^\pm$ plasma created between the radii $r_1 = 3.52\times10^7$ cm and $r_2 = 4.87\times10^8$ cm with a total number of $e^{\pm}$ pairs $N_{e^\pm} = 1.6\times 10^{59}$ and an initial temperature $T = 1.7$ MeV. The theoretically estimated total isotropic energy emitted in the P-GRB is $E_{P-GRB}=1.1\% E_{e^\pm}^{tot}=1.54 \times 10^{52}$ erg, in excellent agreement with the one observed in the first main pulse ($E_{P-GRB}^{obs} \sim 1.5 \times 10^{52}$ erg in $2-700$ keV energy band, see Fig.~\ref{970228_fit_prompt}), as expected due to their identification. After the transparency point at $r_0 = 4.37\times 10^{14}$ cm from the progenitor, the initial Lorentz gamma factor of the fireshell is $\gamma_0 = 199$. On average, during the afterglow peak emission phase we have for the CBM $\langle {\cal R} \rangle = 1.5\times 10^{-7}$ and $\langle n_{cbm} \rangle = 9.5\times 10^{-4}$ particles/cm$^3$. This very low average value for the CBM density is compatible with the observed occurrence of GRB970228 in its host galaxy's halo \citep{sau97,vp,panaitescu06} and it is crucial in explaining the light curve behavior.

The values of $E_{e^\pm}^{tot}$ and $B$ we determined are univocally fixed by two tight constraints. The first one is the total energy emitted by the source all the way up to the latest afterglow phases (i.e. up to $\sim 10^6$ s). The second one is the ratio between the total time-integrated luminosity of the P-GRB and the corresponding one of the whole afterglow (i.e. up to $\sim 10^6$ s). In particular, in GRB970228 such a ratio results to be $\sim 1.1\%$ (see sec.~\ref{fireshell} and Fig. \ref{figX}). However, the P-GRB peak luminosity actually results to be much more intense than the afterglow one (see Fig.~\ref{970228_fit_prompt}). This is due to the very low average value of the CBM density $\langle n_{cbm} \rangle = 9.5\times 10^{-4}$ particles/cm$^3$, which produces a less intense afterglow emission. Since the afterglow total time-integrated luminosity is fixed, such a less intense emission lasts longer than what we would expect for an average density $\langle n_{cbm} \rangle \sim 1$ particles/cm$^3$.

\section{Rescaling the CBM density}\label{rescale}

We present now an explicit example in order to probe the crucial role of the average CBM density in explaining the relative intensities of the P-GRB and of the afterglow peak in GRB970228. We keep fixed the basic parameters of the source, namely the total energy $E_{e^\pm}^{tot}$ and the baryon loading $B$, therefore keeping fixed the P-GRB and the afterglow total time-integrated luminosities. Then we rescale the CBM density profile given in Fig. \ref{mask} by a constant numerical factor in order to raise its average value to the standard one $\langle n_{ism} \rangle = 1$ particle/cm$^3$. We then compute the corresponding light curve, shown in Fig. \ref{picco_n=1}.

We notice a clear enhancement of the afterglow peak luminosity with respect to the P-GRB one in comparison with the fit of the observational data presented in Fig. \ref{970228_fit_prompt}. The two light curves actually crosses at $t_a^d \simeq 1.8\times 10^4$ s since their total time-integrated luminosities must be the same. The GRB ``rescaled'' to $\langle n_{ism} \rangle = 1$ particle/cm$^3$ appears to be totally similar to, e.g., GRB050315 \citep{050315} and GRB991216 \citep{rubr,spectr1,rubr2}.

It is appropriate to emphasize that, although the two underlying CBM density profiles differ by a constant numerical factor, the two afterglow light curves in Fig. \ref{picco_n=1} do not. This is because the absolute value of the CBM density at each point affects in a non-linear way all the following evolution of the fireshell due to the feedback on its dynamics \citep{PowerLaws}. Moreover, the shape of the surfaces of equal arrival time of the photons at the detector (EQTS) is strongly elongated along the line of sight \citep{EQTS_ApJL2}. Therefore photons coming from the same CBM density region are observed over a very long arrival time interval.

\section{Conclusions}\label{concl}

We conclude that GRB970228 is a ``canonical GRB'' with a large value of the baryon loading quite near to the maximum $B \sim 10^{-2}$ (see Fig. \ref{figX}). The difference with e.g. GRB050315 \citep{050315} or GRB991216 \citep{rubr,spectr1,rubr2} is the low average value of the CBM density $\langle n_{cbm} \rangle \sim 10^{-3}$ particles/cm$^3$ which deflates the afterglow peak luminosity. Hence, the predominance of the P-GRB, coincident with the initial spikelike emission, over the afterglow is just apparent: $98.9\%$ of the total time-integrated luminosity is indeed in the afterglow component. Such a low average CBM density is consistent with the occurrence of GRB970228 in the galactic halo of its host galaxy \citep{sau97,vp}, where lower CBM densities have to be expected \citep{panaitescu06}.

We propose GRB970228 as the prototype for the new class of GRBs comprising GRB060614 and the GRBs analyzed by \citet{nb06}. We naturally explain the hardness and the absence of spectral lag in the initial spikelike emission with the physics of the P-GRB originating from the gravitational collapse leading to the black hole formation. The hard-to-soft behavior in the afterglow is also naturally explained by the physics of the relativistic fireshell interacting with the CBM, clearly evidenced in GRB031203 \citep{031203} and in GRB050315 \citep{050315}. Also justified is the applicability of the Amati relation to the sole afterglow component \citep[see][]{amati06,amatiIK}.

This class of GRBs with $z \sim 0.4$ appears to be nearer than the other GRBs detected by \emph{Swift} \citep[$z \sim 2.3$, see][]{guetta06}. This may be explained by the afterglow peak luminosity deflation. The absence of a jet break in those afterglows has been pointed out \citep{campana_short,watson_short}, consistently with our spherically symmetric approach. Their association with non-star-forming host galaxies appears to be consistent with the merging of a compact object binary \citep{barthelmy_short,fox_short}. It is here appropriate, however, to caution on this conclusion, since the association of GRB060614 and GRB970228 with the explosion of massive stars is not excluded \citep{da06,ga00}.

Most of the sources of this class appear indeed not to be related to bright ``Hypernovae'', to be in the outskirts of their host galaxies \citep[][see above]{fox_short} and a consistent fraction of them are in galaxy clusters with CBM densities $\langle n_{cbm} \rangle \sim 10^{-3}$ particles/cm$^3$ \citep[see e.g.][]{la03,ba07}. This suggests a spiraling out binary nature of their progenitor systems \citep{KMG11} made of neutron stars and/or white dwarfs leading to a black hole formation.

It is now essential: 1) to confirm that also GRB060614 conforms to our ``canonical GRB'' scenario, with a deflated afterglow originating in a peculiarly low CBM density (Caito et al. in preparation); 2) to search for a ``genuine'' short GRBs with $B \la 10^{-5}$, possibly GRB050509B \citep[][De Barros et al. in preparation]{ge05}.

\acknowledgements

We thank ICRA-BR and M. Novello for organizing the $1^{st}$ Cesare Lattes meeting where some preliminary results of this analysis have been presented on 2007-02-28 in the tenth anniversary of GRB970228; the Italian Swift Team (supported by ASI Grant I/R/039/04 and partly by the MIUR grant 2005025417) for discussions about GRB060614 and the other ``fake'' short GRBs observed by \emph{Swift}; M. Kramer, F. Frontera and L. Amati for discussions; an anonymous referee for interesting suggestions.

\end{document}